\documentclass[a4paper,11pt]{article}
\usepackage{jinstpub} 
\usepackage[separate-uncertainty]{siunitx}
\usepackage{isotope}
\DeclareSIUnit{\PE}{PE}
\DeclareSIUnit{\SPS}{SPS}
\DeclareSIUnit{\bitpersecond}{bps}
\usepackage{placeins}
\usepackage{sansmath}
\usepackage{multirow}
\DeclareUnicodeCharacter{2212}{−}
\usepackage{lineno}
\compress
\title{Cherenkov Neutrino Telescopes: Recent Progress and Next Steps}

\emailAdd{aya@hepburn.s.chiba-u.ac.jp}
\author{Aya Ishihara}
\affiliation{Dept. of Physics and The International Center for Hadron Astrophysics, Chiba University, Chiba 263-8522, Japan}
\abstract{
Neutrino telescopes provide a unique observational gateway to the high-energy universe, enabling the study of cosmic accelerators and extreme environments that remain inaccessible to the other high-energy messengers. Although they share core detection principles with neutrino experiments in particle physics—such as the observation of Cherenkov radiation—their scientific objectives and operational constraints diverge markedly. This paper reviews the motivations behind astrophysical neutrino detection, outlines key design strategies across various media and deployment environments, and highlights the critical role of neutrino telescopes in the context of multimessenger astronomy. In particular, we emphasize their potential to illuminate the origins of cosmic rays and to probe the mechanisms driving the most energetic phenomena in the universe.
}


\begin{document}

\maketitle
\section{Introduction}

\begin{figure}[htbp]
    \centering
    \includegraphics[width=0.95\textwidth, clip]{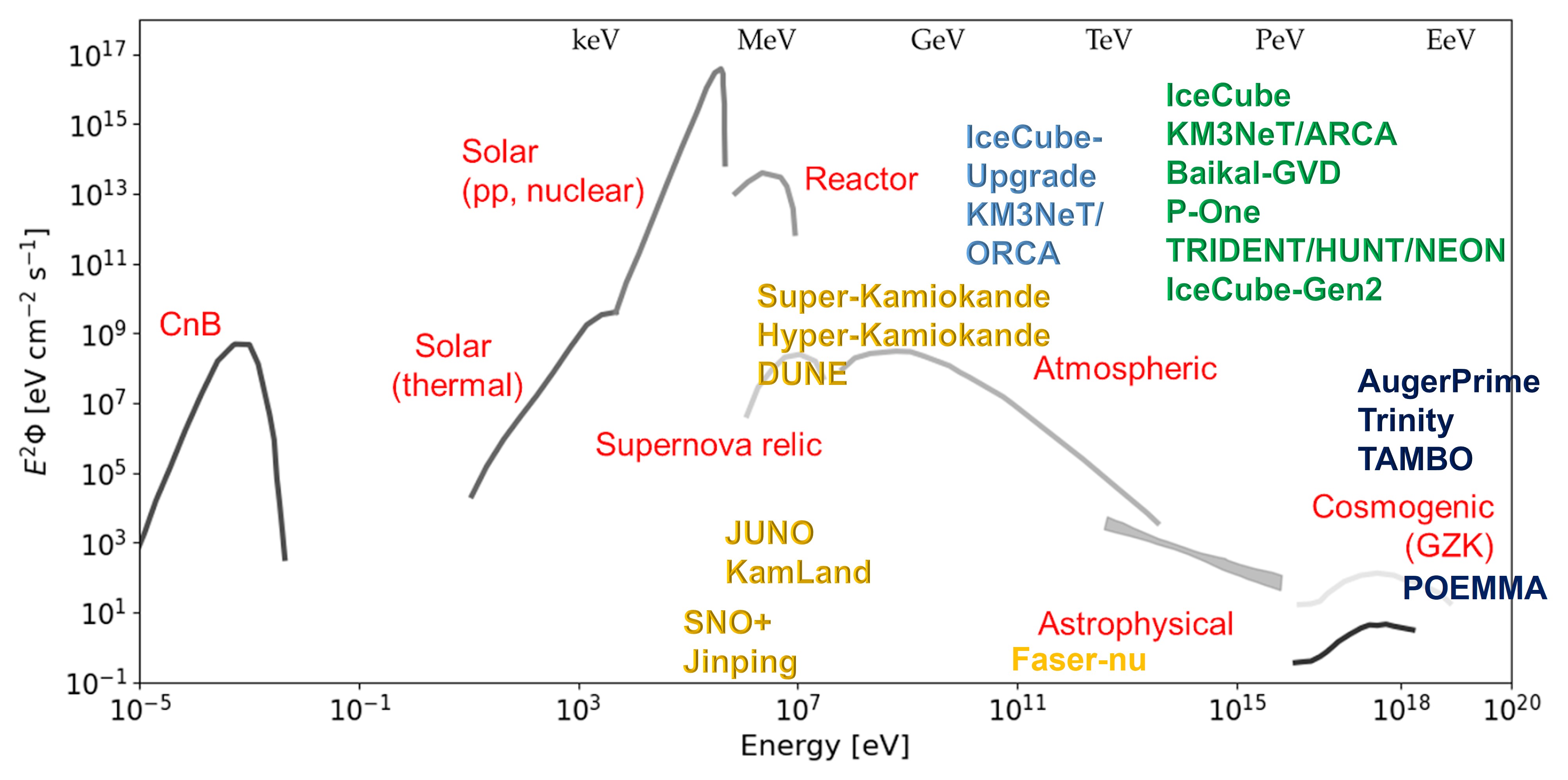}
    \caption{Energy spectra of neutrinos and the corresponding detection ranges. Red labels indicate the neutrino sources included in the flux shown by the gray lines (the black line represents the expected lowest level of the cosmogenic flux). The other labels mark current and future neutrino detectors and neutrino telescopes.}
    \label{fig:overview}
\end{figure}

Neutrinos are among the most abundant particles in the universe, yet they remain remarkably elusive. Therefore, neutrinos went unnoticed—despite being everywhere for a long time. Their weak interaction with matter allows them to traverse vast cosmic distances and dense astrophysical environments virtually unimpeded. As illustrated by the gray lines in Fig.~\ref{fig:overview}, the energy flux of neutrinos spans roughly 25 orders of magnitude in energy. Remarkably, we have already detected neutrinos across 10 of those orders with currently operating detectors, and extensions by currently planned projects promise to cover a few more orders, as indicated in the figure, enabling us to extract valuable information from this naturally occurring flux~\cite{Snowmass_neutrino}.

In particle physics, neutrino detectors have long been used to study fundamental properties of neutrinos, such as their masses and mixing angles, through the observation of neutrino flavor oscillations. These experiments focus on propagation effects over terrestrial baselines.
In contrast, the primary goal of neutrino telescopes is to identify where and how astrophysical neutrinos—and their parent cosmic-ray protons and nuclei—are produced. They are therefore designed to detect naturally occurring neutrinos from cosmic sources, using them as tracers of cosmic accelerators. 
Their unique property—being weakly interacting and electrically neutral—makes neutrinos powerful messengers of high-energy phenomena in the universe, capable of probing regions opaque to photons and of tracing sources where unknown magnetic fields bend the trajectories of charged cosmic-ray particles.

Despite relying on similar fundamental detection principles, neutrino detectors and telescopes differ significantly in design and optimization because their scientific goals are distinct. In particle physics, the focus is on studying propagation effects—such as matter-induced modifications of flavor oscillations—so the primary interest lies in the MeV to GeV energy range. In astroparticle physics, by contrast, the relevant energies extend from tens of TeV, where the observational window opens above the atmospheric background, up to beyond the EeV scale. MeV–GeV neutrinos are also of considerable astrophysical interest in specific cases, as nearby transient sources can produce detectable neutrino emission in this energy band.

\section{Neutrino and multimessenger astronomy}

\begin{figure}[htbp]
    \centering
    \includegraphics[width=0.95\textwidth, clip]{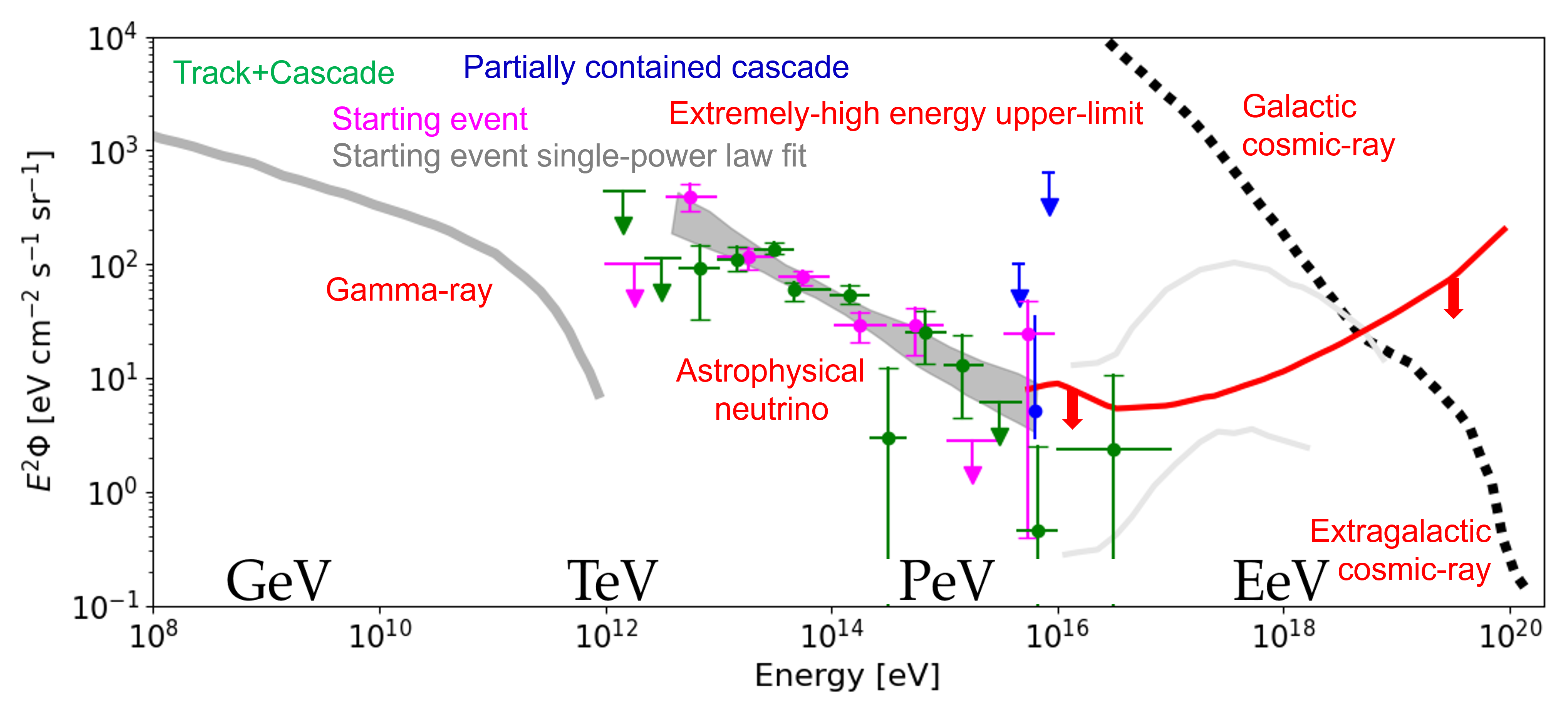}
\caption{Multimessenger spectrum showing diffuse gamma rays, neutrinos, and cosmic rays across energy scales. The gray line indicates the gamma-ray flux, the black dotted line represents the galactic and extragalactic cosmic-ray flux, and the colored markers and the gray band show measured neutrino fluxes~\cite{estes_2024, combinedfit_2025, combinedfit_2025b, glashow_2021}. The red line indicates the upper limit from IceCube’s extremely high-energy neutrino search~\cite{ehe_2025}.}
    \label{fig:gamma_neutrino_cr}
\end{figure}

Neutrino astronomy is motivated by the observation that extragalactic cosmic rays reach energies above  $10^{20}$~eV (Fig.~\ref{fig:gamma_neutrino_cr}). The mechanisms and environments capable of accelerating particles to such extreme energies—and channeling that energy into charged cosmic rays—are of significant scientific interest. However, magnetic fields throughout the universe deflect the trajectories of charged cosmic rays, preventing them from pointing back to their sources. As a result, identifying and understanding the origins of extragalactic cosmic rays has remained a longstanding problem.

The challenges in accessing the high-energy universe are also evident in the multimessenger spectrum shown in Fig.~\ref{fig:gamma_neutrino_cr}: gamma rays above a few hundred TeV interact with background photon fields—most notably the cosmic microwave background (CMB)—leading to electron–positron pair production. As a result, these high-energy gamma rays lose energy and become significantly attenuated over cosmological distances. The interaction length of very-high-energy gamma rays is set by the inverse of the Thomson cross section multiplied by the photon number density, which for the CMB is approximately 400 photons/cm³. Consequently, PeV gamma rays can travel only on the order of $\sim$10 kiloparsecs before undergoing substantial attenuation. 

Neutrinos, by contrast, do not suffer from such limitations. Their weak interaction cross-section allows them to traverse cosmological distances without significant energy loss, making them indispensable probes of the high-energy universe and potential tracers of cosmic-ray origins. Neutrinos fill the notable energy gap between the fluxes of extragalactic cosmic rays and diffuse gamma rays: galactic cosmic rays obscure the extragalactic component below a few EeV, while low-energy background photons prevent gamma rays above a few hundred TeV from propagating over cosmological distances. Neutrinos therefore provide access to regions and energies inaccessible to other messengers and offer a unique means to investigate the connections among extragalactic cosmic rays, gamma rays, and neutrinos.

Thus, a central question for neutrino telescopes is where and how cosmic particles are accelerated. High-energy protons in astrophysical accelerators can interact with ambient photons or matter, producing pions through $p\gamma$ or $pp$ processes. The subsequent decay of these pions generates both gamma rays and neutrinos, linking the two messengers to the same underlying hadronic interaction. Unlike gamma rays and charged cosmic rays, neutrinos are unaffected by magnetic fields and can escape even very dense environments, allowing them to probe even the innermost regions of astrophysical sources. High-energy neutrinos are expected to originate from hadronic interactions in extreme environments, such as active galactic nuclei (AGN), gamma-ray bursts (GRBs), pulsars, and starburst galaxies. They carry information about the energy spectrum, composition, and spatial distribution of their parent cosmic rays, providing a unique window into the physics of these accelerators. Their ability to traverse cosmological distances without attenuation further makes them indispensable messengers of the high-energy universe. Moreover, multimessenger astronomy combines observations of photons, cosmic rays, neutrinos, and gravitational waves to construct a comprehensive picture of astrophysical phenomena. Each messenger provides distinct insights: photons probe electromagnetic processes, cosmic rays provide information on charged particle acceleration, gravitational waves reveal the dynamics of compact objects, and neutrinos uncover hadronic interactions across a wide range of astrophysical environments.

Completed in 2011, the IceCube Neutrino Observatory~\cite{icecube_instrumentation} has been continuously collecting data for over 14 years.
Despite the difficulty of detecting neutrinos due to their extremely low interaction cross-section, IceCube has successfully observed diffuse astrophysical neutrinos in the energy range from $\sim$10~TeV to $\sim$10~PeV, as shown by the green, pink, and blue points and the gray band in Fig.~\ref{fig:gamma_neutrino_cr}. These detections provide a strong motivation for neutrino astronomy and raise fundamental questions about the nature and origin of these cosmic neutrinos.

The observed energy flux of neutrinos is comparable to that of cosmic rays and background gamma rays, suggesting a possible connection between their astrophysical sources. If these messengers originate from the same class of cosmic objects, it implies that the amount of escaping cosmic rays and the amount of neutrinos produced via interactions are of similar magnitude. This balance offers a compelling clue about the physical conditions and particle interactions within these sources. Further insights can be gained by examining the spectral shape of the observed neutrino flux. For instance, neutrinos produced via $p\gamma$ interactions typically exhibit a threshold-like spectrum due to the dominance of the $\Delta$-resonance region in the cross-section. In contrast, neutrinos from $pp$ collisions tend to mirror the parent proton spectrum, resulting in a smoother, power-law distribution. This distinction offers a pathway to identifying the dominant production mechanism based on spectral features. IceCube continues to pursue this goal by analyzing the spectral shape across multiple detection channels. However, current statistics remain insufficient for a conclusive determination. In the era of next-generation neutrino telescopes, we anticipate the ability to distinguish between different source classes and production mechanisms, thereby quantifying the contributions from various types of neutrino-emitting astrophysical objects.
\section{Detection Principles and Medium Properties}

Neutrinos are of profound interest in both particle physics and astrophysics. The development of neutrino detectors and telescopes, however, has followed divergent trajectories, shaped by the differing flux levels of neutrinos across energy ranges—often masked by the overwhelming atmospheric neutrino background—and by the associated engineering constraints. Water Cherenkov detectors, such as Super-Kamiokande, were constructed in the 1980s to study solar, atmospheric, and supernova neutrinos, as well as to search for rare processes such as proton decay. The landmark observation of neutrino oscillations was made possible by these detectors. Dedicated neutrino telescopes—designed to observe astrophysical neutrinos—did not become operational until the 2010s. This temporal gap reflects the formidable challenge of detecting high-energy neutrinos, which requires instrumenting volumes on the order of a cubic kilometer. For context, a cubic kilometer of IceCube corresponds to roughly 16,000 times the volume of Super-Kamiokande and 2,000 times that of Hyper-Kamiokande. Achieving such dimensions has demanded innovations in technology, infrastructure, and cost-effective design.

Although the underlying detection principle—observing Cherenkov radiation from secondary particles—remains consistent, scaling to larger volumes requires a fundamentally different approach. Cost considerations are critical, motivating the use of naturally occurring media such as deep glacial ice or ocean water. These environments allow for vast instrumented volumes but impose stringent requirements on mechanical durability, calibration accuracy, and deployment logistics. Scalability is also a key concern: large-scale detectors are typically constructed in phases, with initial prototypes validating design concepts at sub-kilometer scales. Power- and bandwidth-efficient electronics, modular sensor architectures, and repeatable deployment procedures are essential for successful expansion.

Therefore, a foundational step in the design of a neutrino telescope is the detailed characterization of the optical properties of the naturally occurring detection medium~\cite{ice_southpole_2006}. These properties, particularly the absorption and scattering lengths of UV photons, can differ significantly from those of conventional water or ice. For example, measurements at the KM3NeT site in the Mediterranean Sea~\cite{km3net_tdr} report absorption lengths of approximately 60~m and scattering lengths of 250~m, whereas Lake Baikal exhibits attenuation lengths roughly one-quarter as large~\cite{water_baikal_2024}. Glacial ice provides a stable and optically favorable medium for constructing large-scale detectors, with long absorption lengths and relatively short scattering lengths. 

\begin{figure}[htbp]
    \centering
    \includegraphics[width=0.95\textwidth, clip]{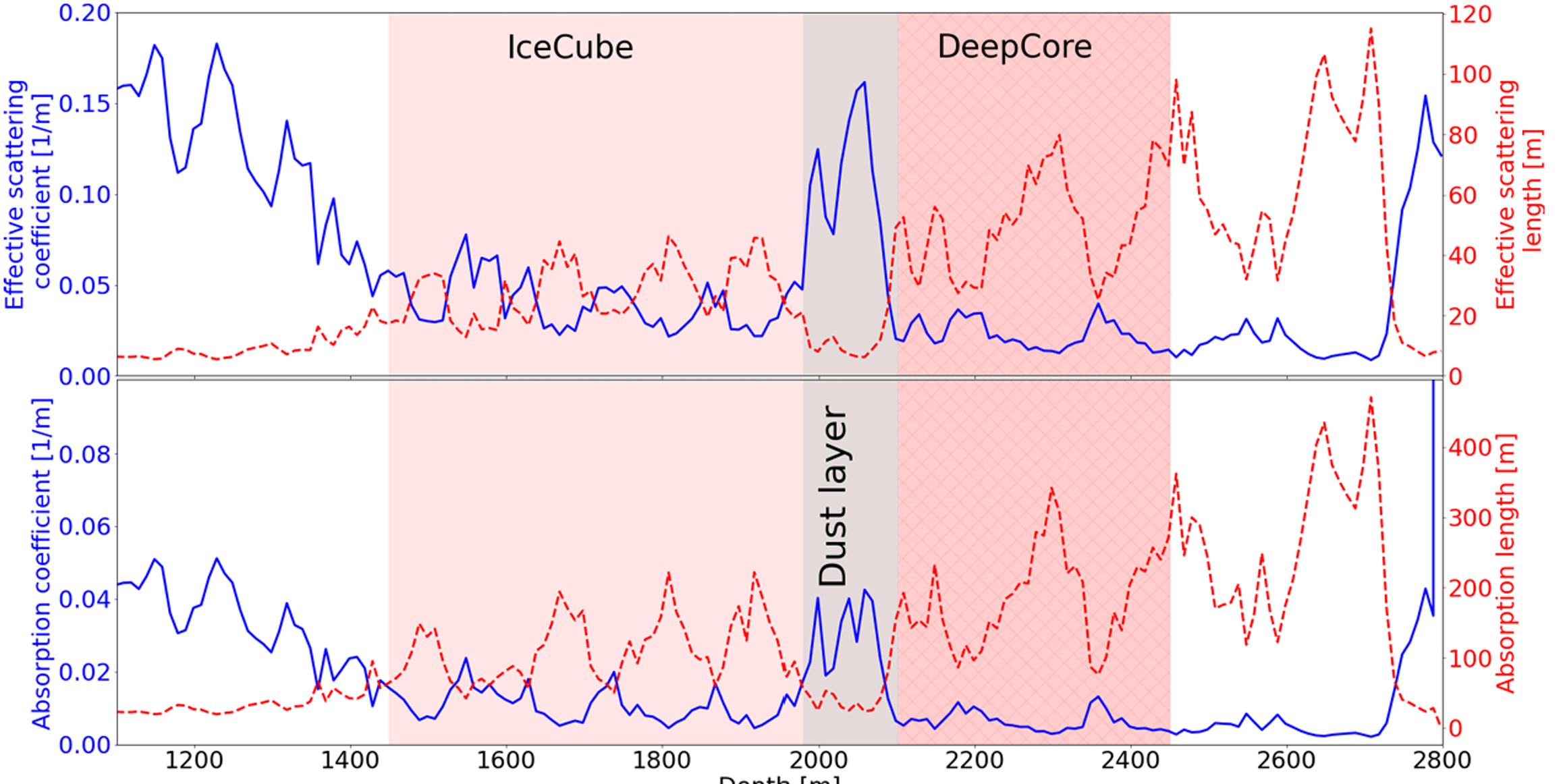}
    \caption{Depth-dependent scattering and absorption properties of South Pole glacial ice. Further details are provided in~\cite{ice_led_2024}.}
    \label{fig:ice_properties}
\end{figure}

Polar ice sheets form through the gradual compression of snow over tens of thousands of years, trapping air between grains and forming bubbles. At depths of a few hundred meters, these bubbles are still visible. Deeper layers undergo a transformation into clathrate hydrates, where air becomes chemically bonded within the ice lattice. At 1300~m depth, the ice appears optically clear, yet each crystal contains nearly 200 times its own volume in trapped air. Although the form of the air changes, its total content remains nearly constant. These changes in microstructure strongly influence optical clarity: at the South Pole, measurements below 1400 m depth show absorption lengths of $\sim$200 m and scattering lengths of $\sim$25 m. The precise values vary with depth according to the ice composition, dust loading, and residual bubble content, as illustrated in Fig.~\ref{fig:ice_properties}~\cite{ice_led_2024}. These models are incorporated into simulation frameworks to accurately reproduce photon propagation. 

Water-based detectors typically require denser sensor arrays to compensate for shorter absorption lengths and extended scattering paths. This configuration ensures sufficient photon collection and timing resolution for reliable event reconstruction. Moreover, angular resolution is generally superior in water-based telescopes compared to their ice-based counterparts. By contrast, ice-based detectors benefit from longer absorption lengths, which allow for wider horizontal spacing between sensors and thereby improve cost efficiency without degrading overall performance.

\section{Deployment Strategies and Calibrations}

As discussed earlier, scalability and repeatable deployment procedures are essential for building detectors at the cubic-kilometer scale, and each observatory has developed a deployment strategy adapted to the specific conditions of its environment. KM3NeT~\cite{km3net_loi_2016}, located in the Mediterranean Sea, relies on a deep-sea method in which each detection unit is coiled around a spherical Launcher of Optical Modules (LOM). The LOM is lowered from a dynamically positioned vessel to the seafloor; once secured in place, a remotely operated vehicle connects interlink cables and triggers the release mechanism. The launcher then rises toward the surface, unrolling the vertical string of optical modules in a controlled ascent. This technique minimizes ship time, permits compact transport, and allows flexible array geometries optimized for directional reconstruction. Baikal-GVD~\cite{baikal_2021} employs a markedly different strategy that leverages the unique winter conditions of Lake Baikal. During the coldest months, the frozen lake surface forms a natural, stable work platform. Access holes are drilled through the ice, and detector strings are lowered and anchored into the water below. The seasonal ice cover facilitates the transport of heavy equipment, simplifies logistical operations, and allows expansion of the array through the sequential installation of additional clusters each winter. 

IceCube, at the South Pole, required a deployment approach optimized for the Antarctic ice sheet. Its construction relied on a high-power hot-water drilling system~\cite{icecube_drill}, capable of melting boreholes roughly 60 cm in diameter down to depths of 2.5 km. In later seasons of the IceCube construction, each hole was drilled in about 30 hours, followed by roughly 10 hours to lower a complete string of 60 optical modules. The water-filled holes refroze over the course of about a week, permanently embedding the sensors within the glacial ice. The hot-water drill system delivered several megawatts of thermal power, and the entire operation was engineered for rapid, repeatable cycles, enabling—in the most efficient periods—the installation of one string per drill camp per day.

Calibration is a foundational component of neutrino telescope operation. Unlike laboratory-based experiments, which benefit from controlled conditions and well-characterized beam parameters, neutrino telescopes operate in vast, naturally occurring media—glacial ice, ocean water, or lake water—where environmental variability and long-term stability present unique challenges. Accurate calibration is essential for understanding detector response and enabling precise reconstruction of neutrino energy, direction, and flavor. The primary objectives of calibration systems in neutrino telescopes include:
\begin{itemize}
    \item \textbf{Geometric calibration:} Determining the precise positions and orientations of optical sensors, which may shift due to environmental forces or deployment tolerances.
    \item \textbf{Timing calibration:} Synchronizing the responses of photomultiplier tubes (PMTs) and associated electronics across the bundles of cables spanning the cubic-kilometer array to ensure precise timing measurements of Cherenkov photons.
    \item \textbf{Optical property mapping:} Measuring absorption and scattering characteristics of the medium to inform photon transport models.
    \item \textbf{Sensor response characterization:} Monitoring gain and noise of individual PMTs to ensure uniform performance across the array.
\end{itemize}

In water-based detectors, calibration must contend with dynamic environmental conditions. In ocean-based telescopes, this complexity is further heightened by currents, temperature gradients, and biological activity, all of which can introduce sensor motion, biofouling, and time-dependent changes in optical properties. KM3NeT addresses these challenges through an integrated calibration system that continuously monitors the geometry and timing of the detector. Acoustic positioning sensors track the three-dimensional locations of detection units with sub-meter precision, while compasses and tiltmeters record the orientation of each optical module to ensure accurate reconstruction of photon arrival directions. Complementing these systems, networks of LED and laser beacons provide timing calibration and probe water transparency, enabling the detector to maintain stable performance despite the ever-changing marine environment.

Baikal-GVD, deployed in Lake Baikal, benefits from comparatively stable freshwater conditions but still requires regular calibration to ensure long-term detector fidelity. Each winter, when the lake surface freezes and forms a natural working platform, teams perform seasonal calibration campaigns that include the deployment of LED flashers, verification of acoustic positioning sensors, and repeated measurements of water clarity. The combination of stable winter access and modular array design enables Baikal-GVD to maintain precise timing and geometry control while gradually expanding its instrumented volume.

IceCube’s calibration framework takes advantage of the long-term stability of Antarctic ice, though optical properties vary with depth due to layers of bubbles and dust. Calibration relies on LED flashers embedded in each optical module to measure timing offsets and local photon propagation, along with laser emitters. These measurements are complemented by dust-logger data collected during drilling~\cite{dustlogger}, which provide a detailed depth profile of dust concentration and inform the optical ice model used in simulations and reconstruction.

\section{The IceCube Upgrade as a Calibration and Testbed Platform}
Following more than a decade of successful operation, IceCube is undergoing a major upgrade to enhance sensitivity to low-energy neutrinos, improve calibration fidelity, and serve as a testbed for next-generation instrumentation for IceCube-Gen2. Construction is scheduled for the 2025/2026 South Pole season, with seven new strings carrying roughly 800 optical sensors and calibration devices deployed near the center of the existing array.

\begin{figure}[htbp]
    \centering
    \includegraphics[width=0.95\textwidth, clip]{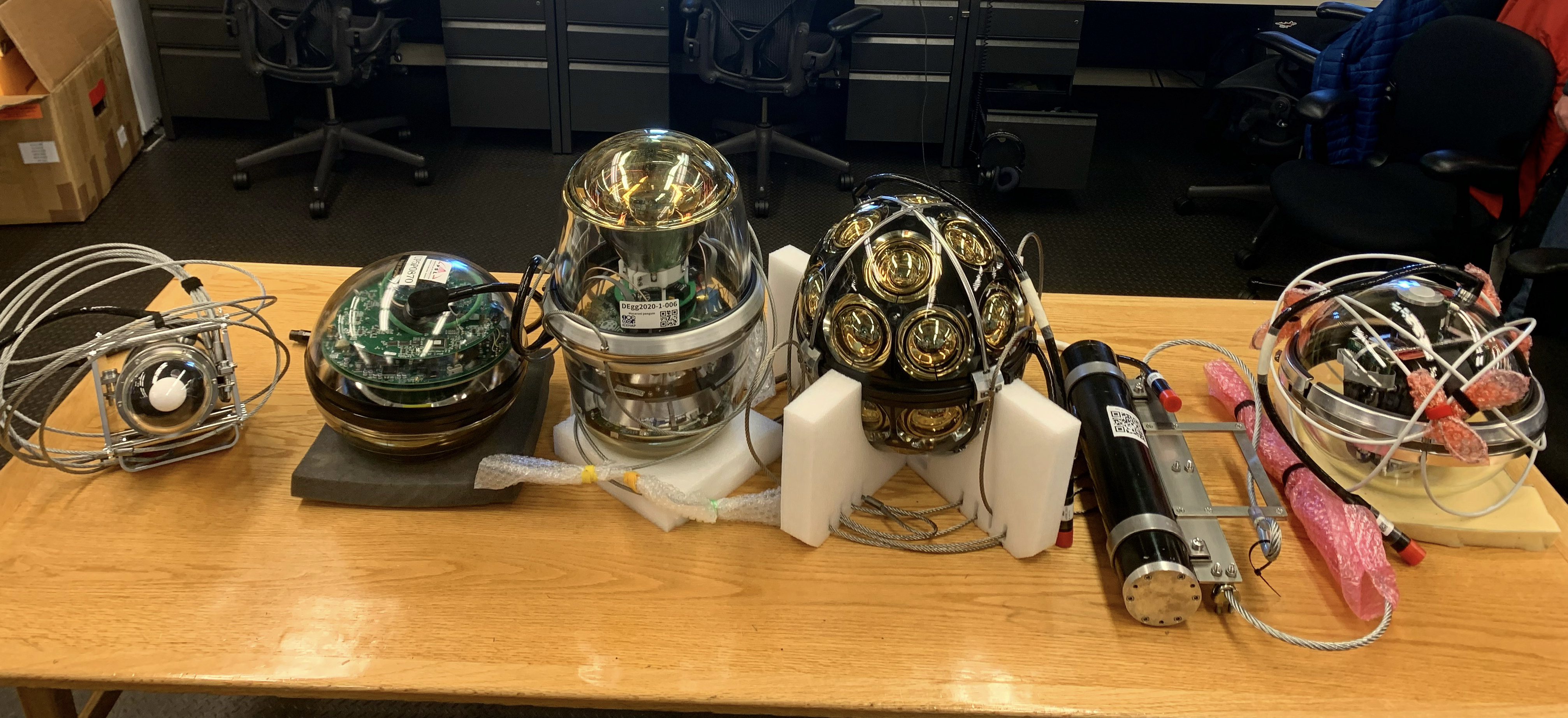}
    \caption{Representative subset of Upgrade modules that had arrived at the Amundsen–Scott South Pole Station by the end of the 2024/2025 season. The two modules on the white stands at the center are a D-Egg and an mDOM, alongside a PDOM on the gray stand.}
    \label{fig:upgrade_modules}
\end{figure}

The deep and densely instrumented region comprises more than 600 optical modules of two new types: D-Eggs~\cite{degg_2023}, containing two high–quantum-efficiency 8-inch PMTs, and mDOMs~\cite{mdom_2025}, housing 24 three-inch PMTs. These modules, arranged at high-density within this central array, provide over an order of magnitude more effective photocathode area per unit volume than DeepCore, enabling improved reconstruction of low-energy events and enhanced directional sensitivity. Separate from this dense region, the shallower subarray (1375–2050 m) supports cross-calibration with the existing detector, detailed characterization of ice properties above the main volume, and testing of next-generation module designs. The deeper subarray (2450–2600 m) extends below the dense region, pushing the limits of the hot-water drilling system and enabling continuous optical characterization of deep-ice as well as deployment of seismometers near the bedrock for geophysical and environmental studies.

To ensure precise calibration of the IceCube array, a comprehensive suite of instruments is being deployed. Isotropic POCAMs emit nanosecond-scale light pulses at multiple wavelengths, with internal monitoring of output intensity, enabling accurate mapping of both local and bulk ice properties. Directional pencil-beam LEDs produce highly collimated pulses ($\sim$1° divergence with 0.1° angular control) to probe anisotropic scattering and absorption over long distances, while embedded laser sources provide additional directional calibration across extended baselines. Optical cameras, both integrated within the new sensor modules and deployed as standalone units, allow direct visual monitoring of ice structure, including bubbles, cracks, and module alignment, throughout both deployment and operation.

The PDOMs are developed based on the legacy IceCube DOMs but incorporate modernized electronics, enabling direct cross-calibration between the original and upgraded systems and allowing systematic separation of PMT-level responses from electronics-level effects. Each PDOM is also equipped with cameras, acoustic receivers, and scintillator-based muon taggers, which provide known, directional calibration sources using atmospheric muons.

The IceCube Upgrade simultaneously serves as a critical testbed for the next-generation IceCube detector, IceCube-Gen2~\cite{gen2_tdr_2023}, which will deploy approximately 10,000 new optical modules. To ensure readiness for large-scale deployment, in-situ performance studies are conducted as part of the Upgrade. These studies are essential for design verification, evaluating how candidate modules perform under real Antarctic conditions. Within the Upgrade, two variants of Gen2 optical modules are being tested. While they differ in mechanical design, they share the same electrical components, allowing direct comparison of housing, integration, and deployment strategies. Insights from both designs are being merged iteratively to inform the final Gen2 architecture, ensuring it is robust, scalable, and practical for full-scale production.

The IceCube Upgrade also evaluates innovative new technologies, including Wavelength-Shifting Optical Modules (WOMs), which absorb UV Cherenkov light and re-emit photons at longer wavelengths for higher PMT efficiency, and Fiber Optical Modules (FOMs), which use wavelength-shifting fibers coupled directly to external PMTs to enhance signal collection and reduce cost per detection volume. These modules are tested under Antarctic conditions to assess durability, performance, and integration with the existing detector infrastructure.
\section{Summary and Outlook}
Neutrino telescopes provide direct access to the high-energy universe, enabling the study of cosmic accelerators and extreme environments beyond the reach of electromagnetic or gravitational signals. While they share the Cherenkov-based detection principle with particle physics neutrino experiments, their astrophysical goals, operational scales, and environmental constraints differ substantially. Neutrino astronomy is motivated by neutrinos' weak interactions and neutrality, allowing them to traverse cosmological distances without attenuation. Observations from IceCube and other detectors have revealed diffuse astrophysical neutrinos in the 10~TeV to 10~PeV range, bridging the gap between gamma rays and ultra-high-energy cosmic rays and suggesting a link between neutrino production and cosmic-ray sources.

Designing neutrino telescopes requires careful consideration of detection media and scalability. Water Cherenkov detectors like Super-Kamiokande established the foundation, but cubic-kilometer-scale telescopes—IceCube, KM3NeT, and Baikal-GVD—necessitated new strategies. Natural media such as glacial ice or deep water provide cost-effective volumes but pose challenges in optical characterization, mechanical durability, and deployment. Ice offers long absorption lengths and short scattering lengths, while water requires denser instrumentation. Deployment strategies are site-specific: IceCube relies on hot-water drilling; KM3NeT uses deep-sea launchers and remotely operated vehicles; Baikal-GVD utilizes the frozen lake surface for seasonal installation. These approaches reflect the ingenuity needed to operate in extreme environments. Calibration underpins accurate event reconstruction, timing synchronization, and medium modeling. IceCube employs LEDs, lasers, and dust loggers for depth-resolved ice characterization, while KM3NeT and Baikal-GVD use acoustic positioning, optical beacons, and environmental sensors. The IceCube Upgrade adds advanced tools—including isotropic flashers, directional beams, cameras, and environmental monitors—alongside new multi-PMT modules, enhancing sensitivity to low-energy events and complex topologies.

Looking ahead, next-generation detectors—IceCube-Gen2, P-ONE~\cite{pone_2020}, TRIDENT~\cite{trident_2023}, and HUNT~\cite{hunt_2025}—are implementing modular, scalable calibration systems and site-optimized deployment strategies. With these innovations, they will enable detailed spectral studies, flavor composition measurements, and multimessenger correlations, opening new windows into the universe’s most energetic processes.

\bibliographystyle{JHEP}
\bibliography{refs}

\end{document}